\documentstyle[aps,prb,preprint]{revtex}
\begin{document}
\draft
\title{Spontaneous dimer order, excitation spectrum, and
quantum phase transitions in  the 
 $J_{1}-J_{2}$ Heisenberg model}
\author{Valeri N. Kotov}
\address{Department of Physics, University of Florida
\\Gainesville, Florida 32611-8440, USA}
\author{Jaan Oitmaa, Oleg Sushkov, and Zheng Weihong}
\address{School of Physics, University of New South Wales,
\\Sydney 2052, Australia}
\maketitle

\begin{abstract}
We overview some recent work and present new results on
 the ground state properties and the spectrum of
 excitations of the two-dimensional frustrated Heisenberg
 antiferromagnet. Spontaneous dimer order is present in the
quantum disordered phase of this model.
 We study the stability and 
analyze the structure of the spectrum, including the two-particle
singlet  excitation branch throughout the disordered phase, as well as
 in the vicinity of the N\'{e}el critical point.
The variation of the dimer order parameter is also given, and it is
 argued that near the critical point  it reflects the presence of   
the low-energy singlet bound state.

\end{abstract}

\vfill\eject

\narrowtext

\section{Introduction}

Many  of the fundamental
 problems in the field of  low-dimensional quantum spin systems
are those related to the nature of the  ground states,
 the spectrum of excitations, and the stability of 
 quantum  phases without long-range order
 (Sachdev 1999). 
 The major interest boost which these topics have enjoyed 
 is certainly  connected to the discovery of the high-T$_{c}$ 
superconductivity.
 The  high-T$_{c}$  materials are doped antiferromagnetic insulators
 where doping destroys the N\'{e}el long-range order, present in
 the parent compounds (Manousakis 1991). 
 It is believed that the antiferromagnetic correlations are intimately
 tied to superconductivity - a connection that emphasizes the need
 for good understanding of the magnetically disordered ground state.
In a proposal, advocated by Anderson (1973, 1987), the ground state  
 is viewed as a coherent superposition of spins paired into singlet bonds,
 which can "resonate" between different configurations. 
 This resonating valence bond (RVB) state is  an example of a true
 spin-liquid - a state that is both translationally and spin-rotationally
 invariant. Depending on the nature of the correlations between the
 valence bonds,  
 different varieties of RVB's have been considered
 (Fradkin 1991). The spectrum of these systems is however far from being
 well understood.
  
A major recent development in the field of  high-T$_{c}$ materials 
was the discovery of 
 stripes - a spatial modulation of the spin and  
charge densities, which seems  to be related to superconductivity
(Tranquada {\it et al.} 1995, 1997). 
In this connection   the study of   
  dimerized,  valence bond solid-like  (VBS) ground states with
spontaneously  broken translational invariance, has become an important issue. 
The stability of  such dimerized phases was recently
investigated in the framework of the $t-J$ model 
 (Sachdev and Vojta 1999a,b, Sushkov 1999, Vojta and Sachdev 1999). 
 Within the approximation schemes used by these authors,  spontaneously
 dimerized ground states were found to be stable in the region of
 doping, relevant to the  high-T$_{c}$ cuprates. More detailed predictions, 
  relevant to experiments, are currently being
 developed (Sachdev and Vojta 1999b).

 Spontaneous dimer order can also occur due to 
spin frustrating
  interactions - this constitutes the subject of the present
work. The basic model which shows  such behavior is
 the two-dimensional $J_{1}-J_{2}$ Heisenberg model on a square lattice.
 There are at least two major reasons why this model holds a special place
in the physics of spin systems: (1.) By itself it is one of the simplest
 models which exhibits quantum transitions between long-range ordered phases
 and a quantum disordered phase - a topic of fundamental interest (Sachdev 1999),
  (2.) Even though the $J_{1}-J_{2}$  model does not contain charge dynamics,
 the understanding of how translational symmetry is broken in a purely
 insulating spin  background  is crucial to the finite doping situation as well.
 Indeed, the  techniques used in the analysis of the $t-J$ model by 
Vojta  and Sachdev (1999) and Sushkov (1999)  have been tested
 first on  purely spin models and rely heavily on the good handling
 of the zero doping case. Finally, due to its simpler nature,
 the $J_{1}-J_{2}$  model allows for 
  a  comprehensive and accurate description
 of the ground state properties  and the spectrum of excitations, 
 which so far is lacking for the  $t-J$ model. 

 The $J_{1}-J_{2}$  model has been discussed in numerous works
 over the last ten years  and some of  the important issues that have been addressed
 are: (1.) How is the N\'{e}el order, present for small frustration ($J_{2}$), 
destroyed as frustration increases, and (2.) Is a quantum disordered phase
 present in a finite window of frustration, and what is the structure of this 
phase. Let us mention several representative papers which discuss these points,
 without attempting to give a comprehensive literature review. 
 Spin-wave calculations, both at the non-interacting level as well as including
 interactions perturbatively in powers of $1/S$ ($S$ is the spin value),
 have found that the magnetization decreases with increasing frustration,
ultimately vanishing at a critical value 
 (Igarashi, 1993). These calculations however cannot predict the structure of the phase
beyond the instability point, or the location of the phase boundary with high
 accuracy, since as the magnetization decreases more and more powers
 of $1/S$ have to be included (strong spin-wave interactions). 
 Exact diagonalization (ED) of clusters as large as $N=36$
 (Schulz {\it et al.} 1994, 1996)
 have found a finite region of quantum disordered (gapped) phase, but have
 failed to determine with certainty what the dominant correlations or
 type of order (e.g. dimer, plaquette, etc.) are dominant in this phase.
 The ED calculations also  suffer from large finite-size corrections,
 especially for strong frustration. 
 An insight into the structure of the disordered phase was  possible
 with  the help of the large $N$ expansion technique (Read and Sachdev 1990, 1991,
 Sachdev and Read 1991). These authors predicted the quantum disordered phase
to be spontaneously dimerized in a particular (columnar) configuration (see figure 1(a)).
 High order dimer series expansions around this 
configuration were performed  (Gelfand {\it et al.} 1989, Gelfand 1990,
Kotov {\it et al.} 1999a),
 all confirming its stability in a window of frustration.
 Thus the  spontaneously dimerized state has emerged as the most
probable candidate for a disordered ground state.
 Let us mention in this connection that in one-dimensional systems 
 the Lieb-Schultz-Mattis (LSM) theorem guarantees that a gapped
phase in a quantum spin system 
 breaks  translational symmetry (Lieb {\it et al.} 1961).
 Extension of the LSM theorem to two-dimensions was proposed (Affleck 1988) 
 but not proven in the most general case. The large $N$ and dimer series
 results however  seem to confirm the validity of the LSM theorem in
 2D as well, including, in fact, the case of finite doping 
 (Sachdev and Vojta 1999b). 
 
The present work is devoted to discussion of several aspects of the
 spontaneously dimerized phase, with particular emphasis on the
 excitation spectrum, including collective modes (Section II),  
and the evolution of the spin-spin correlations  and the
dimer order parameter throughout the disordered phase 
(Section III). Properties near the N\'{e}el ordered - quantum disordered
phase are also discussed.                     
A preliminary account of some of the   results has 
been given by Kotov {\it et al.} (1999a) and Kotov and Sushkov (1999).

\section{Single-particle and collective low-energy excitations in the
 spontaneously dimerized phase.}
 
The $J_{1}-J_{2}$ model is defined via the Hamiltonian:

\begin{equation}
H = J_{1} \sum_{NN}{\bf S}_{i}.{\bf S}_{j} +
  J_{2} \sum_{NNN} {\bf S}_{i}.{\bf S}_{j}, 
\end{equation}
where $J_{1} \geq 0$ 
 is the nearest-neighbor (NN),  and $J_{2} \geq 0$ -
the frustrating diagonal next-nearest-neighbor (NNN) coupling
 on a square lattice (see figure 1(a)).
 The possible ordered phases for small (large) $J_{2}$ 
are shown in  figure 1(b)(figure 1(c)), and are referred to as the N\'{e}el and
collinear phase, respectively. The phase diagram of the model, as determined
by dimer series expansions in the disordered phase (this work, see below), and
 Ising series expansions in the two ordered phases (Oitmaa and Zheng 1996),
 is shown in figure 2. The disordered phase is stable in the interval
 $0.38 \leq (J_{2}/J_{1}) \leq 0.62$, although  there is uncertainly
 in the exact location of the boundaries due to the poorer convergence
 near them. 
 We proceed with a description of the spectrum of excitations and its evolution
in the quantum disordered phase.

\subsection{Triplet spectrum.}

In order to calculate the spectrum we use the "strong-coupling" philosophy,
 i.e. develop perturbative expansions around a "perfect"  
dimer configuration (represented as a direct product of
singlets). The columnar dimerization pattern used in the expansions
is shown in figure 1(a).  Unlike systems where the dimerization
 is explicit, i.e. caused by alternation in the Heisenberg exchange, 
in the $J_{1}-J_{2}$ model there is no  formally small expansion parameter (both
inter- and intra-dimer couplings are of the same order in the relevant
parameter regime). To  achieve maximum reliability of our results we have developed and
 compared two techniques which take into account the inter-dimer interactions. 
 The first one is the dimer series expansion, where  a series 
for the appropriate observable is generated
 in powers of the inter-dimer interaction to high order (typically around 10).
 The generation and resummation of such  a  series as well as
 other types of  linked-cluster expansions involve a lot of technical details
 (Gelfand {\it et al.} 1990). In what follows we will present the results only.
  Our second approach is based on  diagrammatic resummation of selected (infinite) series
which  we find to give the dominant contribution.  
The two approaches are quite similar in spirit, but technically very different,
 thus allowing us to make sure that the (strong) interaction effects are treated
 properly.

In order to get a feeling for the
 types of interactions between the dimers, it is useful to write the
 Hamiltonian, Eq.(1) in terms of operators creating  triplets ($t_{i \alpha}^{\dagger}$,
where $i$ is the dimer site index, and $\alpha=x,y,z$) from 
 the "perfect" dimer configuration. Such a representation
 was discussed by Sachdev and Bhatt (1990), and 
leads to the effective Hamiltonian:            
\begin{equation}
H = H_{2} +  H_{3}  +  H_{4},
\end{equation}
\begin{equation}
H_{2} = \sum_{\bf{k}}
\left\{ A_{\bf{k}} t_{\bf{k}\alpha}^{\dagger}t_{\bf{k}\alpha}
+ \frac{B_{\bf{k}}}{2}\left(t_{\bf{k}\alpha}^{\dagger}
 t_{\bf{-k}\alpha}^{\dagger}
+ \mbox{h.c.}\right) \right \},
\end{equation}
\begin{equation}
H_{3} =  \sum_{\bf{k_{i}}}
 \mbox{R}({\bf k_{1}},{\bf k_{2}})
 \epsilon_{\alpha\beta\gamma}
 t_{\bf{k_{1}}\alpha}^{\dagger} t_{\bf{k_{2}}\beta}^{\dagger}
 t_{\bf{k_{1}+k_{2}}\gamma}
 + \mbox{h.c.},
\end{equation}
\begin{equation}
H_{4} =
   \sum_{\bf{k}_{i}}
\mbox{T}({\bf k_{1}}-{\bf k_{3}})
 (\delta_{\alpha\delta}\delta_{\beta\gamma}-
\delta_{\alpha\beta}\delta_{\gamma\delta})
 t_{\bf{k_{1}}\alpha}^{\dagger}
t_{\bf{k_{2}}\beta}^{\dagger}t_{\bf{k_{3}}\gamma}
t_{\bf{k_{1}}+\bf{k_{2}}-\bf{k_{3}}\delta},
\end{equation}
where:
\begin{equation}
A_{\bf{k}} = J_{1} - \frac{J_{1}}{2}\cos{k_{x}}
+ (J_{1} - J_{2})\cos{k_{y}} - J_{2}\cos{k_{x}}\cos{k_{y}} = 
J_{1} + B_{\bf{k}},
\end{equation}
\begin{equation}
\mbox{T}({\bf p})=
\frac{J_{1}}{4} \cos{p_{x}}
 + \frac{J_{1}+J_{2}}{2}\cos{p_{y}} +
 \frac{J_{2}}{2}\cos{p_{x}}\cos{p_{y}},
\end{equation}
\begin{equation}
\mbox{R}({\bf p},{\bf q}) = - \frac{J_{1}}{4} \sin{p_{x}}
-  \frac{J_{2}}{2} \sin{p_{x}}\cos{p_{y}} - \{p \rightarrow q\}.
\end{equation}
An important point to be made is that the Hamiltonian, Eq.(2) in terms of the triplets
 is exact, i.e. the cubic $\mbox{R}({\bf p},{\bf q})$ 
and the quartic $\mbox{T}({\bf p})$ vertices are the
only ones generated.
However in addition one has to make sure that no double occupancy
on a single site is allowed, i.e. $t_{i \alpha}^{\dagger}t_{i \beta}^{\dagger}=0$,
 which follows from the fact that a triplet is composed of two spins $S=1/2$
 and consequently it is impossible to create higher on-site spins.
 When the dimer series is developed, this constraint is taken into
account explicitly in every order of the perturbative expansion.
 On the other hand if we
choose to treat the interactions $H_{3}$ and $H_{4}$ diagrammatically,
 which is conveniently done in momentum space, it is useful to introduce
an (infinite) on-site repulsion $U$ as an additional vertex in the theory
(Kotov {\it et al.} 1998):

\begin{equation} H_{U} = U \sum_{i}
t_{\alpha i}^{\dagger}t_{\beta i}^{\dagger}t_{\beta i}t_{\alpha i},
 \ \ U \rightarrow \infty.
\end{equation}
Since the interaction is infinite, it has to be treated to infinite
order of perturbation theory, i.e. replaced by an effective  
 scattering vertex $\Gamma^{\alpha\beta,\gamma\delta} ({\bf{k}},\omega)$.
The latter turns out to have the structure
$\Gamma^{\alpha\beta,\gamma\delta}({\bf{k}},\omega) =
 \Gamma({\bf{k}},\omega)(\delta_{\alpha\gamma}
\delta_{\beta\delta} + \delta_{\alpha\delta}\delta_{\beta\gamma})$
 and has the explicit form (Kotov {\it et al.} 1998):

\begin{equation}
[\Gamma({\bf{k}},\omega)]^{-1}=
- \sum_{\bf{q}} \frac{u_{\bf{q}}^{2}
u_{\bf{k}- \bf{q}}^{2}}{\omega - \omega_{\bf{q}} - \omega_{\bf{k}- \bf{q}}}
 + \left\{ \begin{array}{c} u \rightarrow v \\
\omega \rightarrow -\omega \end{array} \right\}.
\end{equation}
Here $u_{\bf{k}}, v_{\bf{k}}$ are Bogoliubov coefficients,
 arising from the diagonalization of  the quadratic part $H_{2}$, Eq.(3), 
 which are given by the standard expression   
 $u_{\bf{k}}^{2}, v_{\bf{k}}^{2} =
\pm 1/2 + A_{\bf{k}}/2\omega_{\bf{k}}$.
We denote by $\omega_{\bf{q}}$ the one-particle dispersion, which
 on a quadratic level is: $\omega_{\bf{q}}= \sqrt{A_{\bf{q}}^{2}-B_{\bf{q}}^{2}}$,
 and then gets renormalized by the interaction terms. 
 Needless to say, the  vertex $\Gamma({\bf{k}},\omega)$ does not replace exactly 
 $H_{U}$, but can be viewed as the best approximation for
 low density of triplets. It is intuitively clear that if the density increases
 the hard-core constraint  becomes harder to satisfy, leading, on a technical
 level, to generation of additional vertices. We have found however that
 the triplet density $N_{T}=\langle t_{\alpha i}^{\dagger}t_{\alpha i} \rangle =
3 \sum_{\bf{q}} v_{\bf{q}}^{2}$ stays around 0.3 throughout the disordered
 phase, thus justifying our approximation Eq.(10).
 Next, the corresponding Dyson equation
with one-loop self-energies arising
 from the three vertices Eqs.(4,5,10) is solved self-consistently for
 the triplet spectrum $\omega({\bf k})$.
 Explicit formulas and the corresponding diagrams 
for  Hamiltonians with similar structure to Eq.(2) 
 can be found, e.g. in the works by two of us and co-authors (Kotov {\it et al.} 1999b,
Shevchenko {\it et al.} 1999). Here we only present the results. 
 Let us note that, as pointed out by
 Chubukov  and Jolicoeur (1991), the interaction effects are responsible
 for creating a finite window of frustration where the gap is non-zero, i.e.
 the dimer phase is stable\footnote{If one neglects the interactions
Eqs.(4,5,10), the region of stability is limited to the point
 $J_{2}/J_{1}=1/2$ only.}. 

In figure 3 the spectrum calculated by the dimer series expansion (order 8)
 is plotted. As frustration increases, the minimum of the dispersion shifts
from the N\'{e}el ordering wave vector ${\bf Q}_{AF}$ to
the collinear ordering  ${\bf Q}_{COL}$. We work in the Brillouin zone of the
dimerized lattice (doubled  unit cell in the x-direction), where
${\bf Q}_{AF}=(0,\pi)$ and 
 ${\bf Q}_{COL}=(0,0)$. 
The convergence of the dimer series is quite poor in certain regions of
${\bf k}$ space which we attribute to decay, caused by 
an overlap with the two-particle
continuum (seen clearly in figure 4).
The evolution of the gaps at the two ordering wave-vectors
is shown  in figure 2. The points where the gap vanishes
 are the two quantum critical points where  transitions take place into
 the corresponding ordered phases.
We estimate the locations of these points, within errorbar, to be 
$(J_{2}/J_{1})_{c}^{(1)} \approx 0.38$ and $(J_{2}/J_{1})_{c}^{(2)} \approx 0.62$.

 In figure 4 we present a  comparison  between our diagrammatic and
dimer series results for the triplet spectrum for a fixed value of frustration
$J_{2}/J_{1}=0.4$.  We find that the agreement is excellent.
 This figure also shows that the error bars on the dimer series curve
 are the largest (i.e. the series does not converge well)
 in the ${\bf k}$ interval where the dispersion enters
the scattering continuum, as discussed in the previous paragraph.
 One can also compare the gaps at ${\bf Q}_{AF}$ calculated by the two methods
(Kotov {\it et al.} 1999a). Both approaches  produce an almost linear variation
 of  $\omega({\bf Q}_{AF})$  in the disordered phase, and give very similar
 values for $(J_{2}/J_{1})_{c}^{(1)}$, although generally the diagrammatic method
 gives larger gap values compared to the dimer series.
 We believe that the combination of the two methods  leads to a very accurate
 description of the spectrum in the quantum disordered phase.

\subsection{Singlet bound state spectrum.}

We now turn to the description of the spectrum of collective
 two-particle excitations with spin $S=0$. Our motivation for the study of this branch of
the spectrum is two-fold: (1.) Quasi one-dimensional systems, such as spin
chains and ladders (Sushkov and Kotov 1998, Shevchenko {\it et al.} 1999)
 were found to have  well
 defined singlet bound states. Frustration was also generally found to increase the binding
energy of these modes.  Collective states in frustrated 2D systems however have not
 been studied. 
 (2.) A gapped spinless collective mode, generated 
non-perturbatively via instanton effects, appears in the  large-N field theory
solution (Read and Sachdev 1990). Moreover, the energy scale (gap) of this mode
 determines the variation of the dimer order parameter near the N\'{e}el critical
point $(J_{2}/J_{1})_{c}^{(1)}$. It is important to explore this non-trivial
connection from the dimer expansion point  of view,  which takes into account
fluctuation effects differently from the large-N approach, and is expected
 to perform much better numerically.

The wave function of two triplets combined into a state with $S=0$
 is given by ({\bf Q} is the total momentum of the pair):
\begin{equation}
|\Psi_{{\bf Q}}\rangle= \sum_{\bf q}\Psi
({\bf q},{\bf Q})t_{\alpha,{\bf Q}/2 + {\bf q}}^{\dagger}
t_{\alpha,{\bf Q}/2 - {\bf q}}^{\dagger}|0\rangle.
\end{equation}
The attraction between the triplets is mainly due to the two-particle
 scattering vertex $\mbox{T}({\bf k})$, relative to which the (second-order)
contribution of $\mbox{R}({\bf k_{1}},{\bf k_{2}})$ turns out to be quite
small and will be neglected from now on for simplicity. 
Diagrammatically the mutual scattering of two quasiparticles
is shown in figure 5. The equation determining the
 bound state energy $E_{\bf Q}^{S}$ reads:
\begin{equation}
\label{BS}
\left [E_{\bf Q}^{S}-\omega_{{\bf Q/2}+{\bf q}}-\omega_{{\bf Q/2}- {\bf q}}
\right ]\psi({\bf q}, {\bf Q})=
\sum_{\bf p}M^{S}({\bf p}, {\bf q}, {\bf Q})\Psi({\bf p}, {\bf Q}), 
\end{equation}
where
\begin{equation}
M^{S}({\bf p},   {\bf q}, {\bf Q})= \left \{-2[\mbox{T}({\bf p}-{\bf q}) +
 \mbox{T}({\bf p}+{\bf q})] + U \right \}.
\end{equation}
The hard-core repulsion  ($U \rightarrow \infty$) has to be also taken
into account by imposing the following condition
 via a Lagrange multiplier:
\begin{equation}
\sum_{\bf q} \psi({\bf q}, {\bf Q}) = 0.
\end{equation}

In figure 4 we present our results for  the bound state spectrum $E_{\bf k}^{S}$
at $J_{2}/J_{1}=0.4$, relatively to the two-particle scattering continuum.
 The bound state has a finite binding energy $\epsilon({\bf k})$, defined as
 $\epsilon({\bf k}) =
[ \mbox{(Lower edge of continuum)}({\bf k})
- E_{\bf k}^{S} ]$, throughout the Brillouin zone, with the exception of the
point ${\bf Q}_{AF}$ and its vicinity. The binding for ${\bf k}=(0,0)$ is
 quite small due to the closeness of the the N\'{e}el critical point
 (see discussion below).  The strongest binding takes place
 for ${\bf k}=(\pi,\pi/2)$,  where $\epsilon(\pi,\pi/2) \approx J_{1}$. 

In figure 6 we present the evolution of the bound state singlet gap for 
 ${\bf k}=(0,0)$, $E^{S}(0,0)$  which will be of particular interest to us in the next
 section. This gap is larger that the one-particle triplet gap
$\omega({\bf Q}_{AF}) \equiv \Delta$,  and appears  
to go to zero (within our resolution), along with $\Delta$, 
as the N\'{e}el transition point  $(J_{2}/J_{1})_{c}^{(1)} \approx 0.38$
is approached. We must note that our treatment breaks down at the
 very vicinity of  $(J_{2}/J_{1})_{c}^{(1)}$ since 
divergences in our diagrams start appearing 
for  $\Delta \rightarrow 0$. Such divergences can presumably be summed
by renormalization group techniques, which we have not attempted in the present 
work. Therefore the  values of the gaps for $J_{2}/J_{1}=0.38$, shown in
 figure 6 should be viewed as being practically zero within the 
error of our calculation at that point.
 It is clear that since  for  ${\bf k}=(0,0)$ the binding energy
 is given by
 $\epsilon(0,0)=2\Delta-E^{S}(0,0)$, it must approach zero
  at  the critical point   $(J_{2}/J_{1})_{c}^{(1)}$. 
 The full evolution of $\epsilon(0,0)$ is shown in figure 6.
  Around the middle of the quantum disordered phase $J_{2}/J_{1} \approx 0.5$
 the binding is large ($\epsilon(0,0) \approx 0.4 J_{1}$), then it shows
am almost linear decrease, and for $J_{2}/J_{1} < 0.41$  becomes smaller 
than the  accuracy of our calculation. 
Alternatively, in the regime $\Delta \rightarrow 0$ one can study
the asymptotics of  $\epsilon(0,0)$   analytically
 (Kotov and Sushkov 1999) with the  result:          

\begin{equation}
\label{binding}
\epsilon({\bf k}={\bf 0}) \sim  \exp \left ( {-\frac{C_{1}}{\Delta}} \right ),
\  \Delta \ll 1 \ \Longrightarrow \ R \sim \frac{1}{\sqrt{\epsilon \Delta}} 
\sim \sqrt{\xi} \exp{(C_{2} \xi)},
 \  \xi \sim \frac{1}{\Delta} \gg 1.
\end{equation}
Here we have also shown the asymptotic behavior of the
 radius $R$ of the bound state (defined as
the spatial extent of the wave function $\psi({\bf q}, {\bf Q}=0)$). 
All energies are measured in units of $J_{1}$ and $C_{1},C_{2}$ are constants
of order unity. Thus we can see that $\epsilon(0,0)$ vanishes
 exponentially fast as a function of the triplet gap, implying an
 exponentially large radius of the composite state.
 This behavior is certainly not captured well numerically in the vicinity
 of the critical point.

The properties of the spectrum 
 near the spontaneously dimerized-N\'{e}el critical point
 are quite peculiar. In addition to the triplet gap $\Delta$,
 the  singlet (bound state)  gap    goes to zero as well. However the spectral
 weight of the composite singlet, which is determined by
 its binding energy (or size), approaches zero exponentially fast near
the transition point\footnote{This is in contrast to the spectral weight
of the triplet, which stays finite at the transition.}.
 For this reason we do not expect that the singlet mode can affect the
 critical dynamics of the triplets and change the $O(3)$ universality
class, associated with such a transition. To compare our results
with 
 the large-N approach we notice that the latter
 also predicts two large length scales near
the critical pont, corresponding to the small gaps of a  triplet and a  spinless
collective mode (Read and Sachdev 1990).
 One of the scales was found to be a (large) power of the other
one 
 and it was argued (Chubukov {\it et al.} 1994) that  this is characteristic
of a "dangerously irrelevant" coupling, i.e.   
 a coupling  which is irrelevant at the critical fixed point, but relevant in the phase,
 from which the transition is approached.
  Thus the overall structure of the spectrum as well as the appearance
 of the additional singlet mode  and its
 effect  
on the critical properties  seem to be quite similar in ours and the
 large-N field theory approach. We must notice however that the  construction of
the spectrum of excitations (and consequently  all associated details) are
  very different in the two approaches
 and  comparison has not been attempted.

\section{Dimer order parameter.}

In this section we discuss the behavior  of the dimer order parameter (DOP), 
 mainly focusing on two issues: (1.) How does the DOP
 reflect the structure of the spectrum found in the previous section?
(2.) Does the DOP go to zero at the N\'{e}el critical point, i.e.
is spontaneous dimerization possible in the ordered phase?
 Even though we can not give completely definite answers to these
 questions, we believe we can shed some light on the issues.
 The large-N approach gives the  following prediction for the
 DOP near the critical point (Read and Sachdev 1990):
 $\mbox{DOP} \sim \Delta^{\mbox{const.N}}, \Delta 
\rightarrow 0$, where $N \gg 1$ is the parameter of the $1/N$ expansion
($N=1$ is the physical limit),
 and the constant is of order unity. Therefore a very fast (even singular)
 vanishing  of the DOP  at the critical point is expected in this scenario.
However  other works (Gelfand {\it et al.} 1989, Sachdev and Bhatt 1990), 
 which are similar in treatment to  ours,  have not found  
 any substantial decrease of the DOP near the transitions point, thus
 leaving the issue still open.  

Let us start by examining  the spin-spin correlation functions, defined
 as:
$\rho_{x}=\langle {\bf S}_{2}.{\bf S}_{3} \rangle, \ 
\rho_{y}=\langle {\bf S}_{1}.{\bf S}_{5} \rangle, \
\rho=\langle {\bf S}_{1}.{\bf S}_{2} \rangle$,
where the spin indexes refer to the sites, as numbered in figure 1.
 In figure 7 we plot the variation of these quantities obtained by
dimer series expansions to order nine. The strength of the inter-dimer
 correlations in the $x$ direction is weaker than in the
$y$ direction in most of the disordered phase, with the exception
of the regions around the two critical points. Therefore the 
 spontaneously dimerized phase can be viewed as a system of
 weakly coupled two-chain ladders (the chains running in the $y$ direction).
 This interpretation was  pointed out by Singh {\it et al.} (1999).

 The dimer order parameters in the two directions are defined as:
$D_{x}=\rho_{x} - \rho, \ D_{y}=\rho_{y} - \rho$, and their variation is shown
 in figure 8. Unlike the previously reported dimer series calculations 
 (Gelfand {\it et al.} 1989), our results  show a substantial decrease
 of both order parameters near the N\'{e}el critical point. 
 In fact $D_{y}$ appears to be zero in  a wide region around this point,
 up to $J_{2}/J_{1} \approx 0.45$.
 The  difference between ours and the previous results is most likely
 due to  the longer series that we have obtained, and consequently
 a better treatment of the inter-dimer interactions.
 In figure 8 we present also our diagrammatic result for $D_{x}$
 which clearly shows a very small variation ($D_{y}$ has similar
behavior). This is quite puzzling, as the dimer series and the
diagrammatic approach give very similar results for the triplet spectrum.
 We have identified the source of this discrepancy to be in the
 very small variation of the in-dimer spin-spin correlation $\rho$
 in the diagrammatic calculation. Figure 9  shows this quantity, calculated 
by the two 
 methods. Indeed, a substantial difference is found  in the vicinity of
 the critical point. An important effect, leading to an explanation of
 this difference was discussed by Kotov {\it et al.} (1999a).
 Let us elaborate more on that point. One can easily see that
 the correlator $\rho$ is  related to the density of
 quasiparticles $N_{T}$ via:
 
\begin{equation}
\rho = -\frac{3}{4} + N_{T}, \ \ 
 N_{T} = \langle t_{\alpha i}^{\dagger}t_{\alpha i} \rangle =
3 \sum_{\bf{q}} v_{\bf{q}}^{2}.
\end{equation}
From figure 9 the dimer series estimates $N_{T} \approx 0.43$ at the critical
point, to be compared with the smaller diagrammatic value $N_{T} \approx 0.3$. 
The Bogoliubov coefficient $v_{\bf{q}}$ is determined by the strength
 of quantum fluctuations $B_{\bf{q}}$ (Eq.(3)), as can be formally
 seen from its definition (see text following Eq.(10)).
 A sharp increase in $B_{\bf{q}}$ can be due to  a strong mixing
 between the ground state (singlet) and a nearby low-lying singlet state. 
 As discussed in the previous section the two-particle singlet bound state
 has a very low energy near the N\'{e}el critical point  and therefore
  can influence substantially the quantum fluctuations.
Technically this effect is given by the diagram for the anomalous
 Green's function (which renormalizes $B_{\bf{q}}$), shown in the
inset of figure 9.
 In order to demonstrate the influence of the bound state on $B_{\bf{q}}$
 let us consider a simplified form of the two-particle
 interaction $\mbox{T}({\bf k})$ and assume that this is the only
term contributing to the formation
 of the bound state (i.e. neglect the $U$ and $R$ contributions).
Choose 
$\mbox{T}({\bf k}) = (J_{1}/4) \cos{k_{x}}$ (i.e. only the $x$
part of the interaction), which leads to
  a separable kernel $M^{S}({\bf p}, {\bf q}) = 
-J_{1}\cos{p_{x}}\cos{q_{x}}$ (Eq.(13)), and thus allows us to
 write the solution in closed form. In this simple case the
 diagram in figure 9 leads to the following renormalization:

\begin{equation}
B_{\bf{k}} \rightarrow B_{\bf{k}} -\frac{J_{1} \cos{k_{x}}
\sum_{\bf{p}}u_{\bf{p}}v_{\bf{p}}\cos{p_{x}}}{1-(J_{1}/2)\sum_{\bf{q}}
(\cos^{2}{q_{x}}/ \omega_{\bf{q}})}.
\end{equation}
The expression in the numerator is the
first order self-energy contribution, which more generally is:
$\Sigma_{4,A}({\bf{k}})
 = -4\sum_{\bf{q}} {\mbox T}({\bf{q}} +
{\bf{k}})v_{\bf{q}}u_{\bf{q}}$, while the denominator is 
accumulated from the resummation of the series, shown in figure 5.
Eq.(17) is to be compared with the equation for the energy of the
 singlet bound state
 (Eq.(12)),  which in this case is: 

\begin{equation}
1 = -J_{1}\sum_{\bf{q}}\frac{\cos^{2}{q_{x}}}{E_{\bf Q}^{S}-
\omega_{{\bf Q/2}+{\bf q}}-\omega_{{\bf Q/2}- {\bf q}}}.
\end{equation}
Setting ${\bf Q}=(0,0)$ in Eq.(18) one can see that as
the energy of the bound state $E_{{\bf Q}=(0,0)}^{S}$ decreases,
the correction to $B_{\bf{k}}$ increases
 (and in fact diverges in the limit $E_{{\bf Q}=(0,0)}^{S} \rightarrow 0$), leading also 
 to an increase in the density $N_{T}$.
 
Although we have not performed a fully self-consistent calculation of the
effect discussed above, it is quite clear that  the low-energy singlet
bound state is mainly responsible for the increase of the quantum fluctuations
 near the N\'{e}el critical point. For comparison, in
  the large-N theory of Read and Sachdev (1990)     
 there is a direct relationship between the gap in the singlet spectrum and the dimer
 order parameter, leading to the fast vanishing of the latter.
 Even though our dimer series results are consistent (within errorbar) with
 vanishing of the dimer order parameter, we certainly can not determine 
 its critical behavior. 
 
\section{Summary and outlook.}
To summarize our main results:

(1.) We have calculated  the one-particle triplet as well as the
collective  two-particle singlet excitations.
 Both branches of the spectrum were found to be stable (i.e. gapped)
 in the spontaneously dimerized phase. 
 
(2.) The singlet bound state mode reflects  the presence of 
 spontaneous dimer order in the system. The singlet gap ({\bf k}=(0,0)) vanishes at the
N\'{e}el quantum critical point. 
 
(3.) The spontaneous dimer order parameters ($D_{x}$ and $D_{y}$) vanish
 at the N\'{e}el quantum critical point. We have presented arguments
 that the disappearance of the dimer order is related to presence
of the low-energy (ultimately gapless) singlet collective mode.

(4.) The collective singlet mode does not influence the quantum critical
 dynamics of the triplets, since its spectral weight vanishes exponentially
 fast at the critical point. Thus the $O(3)$ universality class  describes
 the transition.

\noindent
The above conclusions are consistent with the large-N field theory
 predictions for this model (Read and Sachdev 1990).

So far we have  not analyzed the region of the phase diagram near the transition to
 the collinear phase, $(J_{2}/J_{1})_{c}^{(2)} \approx 0.62$.
 Based on the analysis of Section III it would be tempting to take as a starting
 point of the analysis  a system of weakly interacting spin ladders. This description
 certainly appears to be well justified in the neighborhood of the point $J_{2}/J_{1}=0.5$.
 The Heisenberg model on a frustrated spin ladder  has been previously
 analyzed (Zheng {\it et al.} 1998, Kotov  {\it et al.} 1999b) and a fairly
good understanding of the various phases was achieved. 
 However it is not clear how far this analogy can be used, since
 near $(J_{2}/J_{1})_{c}^{(2)}$ the strength of the inter-dimer correlations
 in the two spatial directions is practically the same, thus making the single ladder 
  problem not representative of the full two-dimensional one. 
 We also note that the quantum transition near   $(J_{2}/J_{1})_{c}^{(2)}$
 appears to be of first order (Singh {\it et al.} 1999).

Other possible types of spontaneous order, such as the plaquette one, have also
been discussed in the literature. Two works have favored this order as a good
 candidate for a ground state in the quantum disordered regime  
(Zhitomirsky and Ueda 1996, Capriotti and Sorella 1999), whereas a recent
 work based on plaquette series expansions has found this configuration 
 to be unstable (Singh {\it et al.} 1999), thus presenting strong evidence
 that plaquette order does not take place in the $J_{1}-J_{2}$ model.

Finally we mention two frustrated spin systems that could be analyzed with
 techniques, similar to the ones used in this work.

(A.) The isotropic two-dimensional Heisenberg model on a triangular lattice
 is believed to be long-range ordered with a non-collinear
 spin arrangement, but with a very small magnetization.
 A small variation of the couplings however stabilizes a 
spontaneously dimerized phase in this system (Zheng {\it et al.} 1999).
 Since the quantum transition between a non-collinear and a quantum disordered 
 phase is expected to be in the  $O(4)$ universality class (Chubukov 1991, Azaria
and Delamotte 1994), the properties of  singlet modes in such
a system are of particular interest. A singlet mode, in addition
 to the triplet one,  is expected to
 become gapless and relevant at the transition point.
 This scenario
 is very different from the one discussed in the present work.

(B.) The Heisenberg model on a Kagom\'{e} lattice was found to have
 very peculiar properties, such as a gap to triplet excitations (i.e.
 a disordered ground state),   with many  low-energy singlet excitations in
 the gap (Waldtmann {\it et al.} 1998, Mila 1998).
 This structure of the spectrum suggests that a simple
 spontaneous dimerization pattern (of the type discussed in this work)
 is highly unlikely. A more probable scenario is a larger "dominant"
 cluster with strong resonance between the dimers. It remains to be
 seen  whether translational invariance is broken in this system, or it
 represents an example of a true spin liquid.

\acknowledgements

 We would like to thank  S. Sachdev, R. Singh, N. Read,
C. Lhuillier, F. Mila, O. Starykh, M. Vojta, C. Biagini, and A. Chernyshev
for stimulating discussions on these and related topics.
V.N.K. was supported by NSF grant DMR9357474 at the University
 of Florida and is grateful to S. Hershfield for his
 encouragement and interest in the work.
J.O., O.S. and Z.W. acknowledge the support of the Australian
 Research Council.

\begin{figure}
\caption{(a) Square-lattice J$_{1}$-J$_{2}$ model.
 The ovals represent spins bound into singlets, forming
the columnar dimer configuration. (b) N\'{e}el ordered state
 with wave-vector ${\bf Q}_{AF}=(\pi,\pi)$. (c) Collinear state
 with ${\bf Q}_{COL}=(\pi,0)$.
\label{fig1}}
\end{figure}

\begin{figure}
\caption{Phase diagram of the model, as determined by high-order
series expansions. Ising expansions to order 10 (9) for the magnetization
 $M$ were used in the N\'{e}el (collinear) phase. The quantum disordered phase
 is characterized by a gap in the triplet spectrum $\omega({\bf k})$,
 determined by the dimer series expansion to order 8. The series
 were summed by using standard techniques.
\label{fig2}}
\end{figure}

\begin{figure}
\caption{Triplet excitation spectrum for two values of
 frustration in the quantum disordered phase.
 Notice that we work in the Brillouin zone of the dimerized lattice,
 where  ${\bf Q}_{AF}=(0,\pi)$ and ${\bf Q}_{COL}=(0,0)$.
\label{fig3}}
\end{figure}

\begin{figure}
\caption{Excitation spectrum, including the two-particle singlet 
 collective mode, for a value
 of frustration close to the transition point into the N\'{e}el phase.
The solid and the dot-dashed lines are the triplet spectrum calculated
 by the dimer series expansion and  diagrammatically, respectively.
The long-dashed line represents the spectrum of the singlet bound state,
 and the shaded region is the two-particle scattering continuum.
\label{fig4}}
\end{figure}

\begin{figure}
\caption{Equation for the two-particle scattering amplitude
(filled square), 
whose poles 
 determine the energies of  the bound states.
\label{fig5}}
\end{figure}

\begin{figure}
\caption{Bound state 
singlet (dashed line) and triplet (solid line through filled squares)
 gaps, calculated diagrammatically. Opens squares represent the
 calculated singlet binding energy, and the solid line is the  fit,
based on 
 the asymptotic formula Eq.(15). 
\label{fig6}}
\end{figure}

\begin{figure}
\caption{Nearest-neighbor spin-spin correlation functions
inside a dimer ($\rho$) and  in the $x$ ($\rho_{x}$) and $y$ ($\rho_{y}$) directions
 between dimers (see text for definitions). All results are obtained by the
 dimer series expansion.
\label{fig7}}
\end{figure}

\begin{figure}
\caption{The dimer order parameters $D_{x}$ and $D_{y}$
 in the two directions, calculated by
 the dimer series expansion (points, connected by solid lines).
For comparison, the diagrammatic result for $D_{x}$ is also plotted
 (dashed line). 
\label{fig8}}
\end{figure}

\begin{figure}
\caption{The in-dimer spin-spin correlation function $\rho$ (see text
for definition). Comparison is made between the dimer series and the
 diagrammatic result, which does not take into account self-consistently
the diagram, shown in the inset. The filled square in the diagram represents
the effective two-particle interaction from figure 5.
\label{fig9}}
\end{figure}

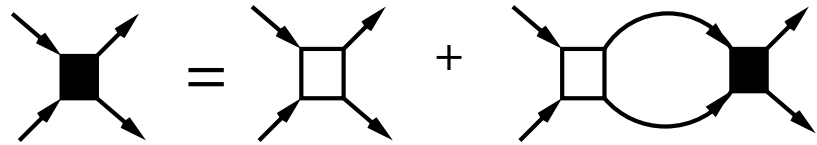
\end{document}